# Broadband ultra-thin acoustic metasurface absorber

Yifan Zhu[*], Krupali Donda, Shiwang Fan, Liyun Cao, and Badreddine Assouar[*]

*Institut Jean Lamour, CNRS, Université de Lorraine, Nancy, France.*

**Abstract**

We theoretically and experimentally propose two designs of broadband low-frequency acoustic metasurface absorbers (Sample I/Sample II) for the frequency ranges of 458Hz~968Hz and 231Hz~491Hz (larger than 1 octave), with absorption larger than 0.8, and having the ultra-thin thickness of 5.2cm and 10.4cm respectively ($\lambda/15$ for the lowest working frequency and $\lambda/7.5$ for the highest frequency). The designed supercell consists of 16 different unit cells corresponding to 16 eigen frequencies for resonant absorptions. The coupling of multiple resonances leads to broadband absorption effect in the full range of the targeted frequency spectrum. In particular, we propose to combine gradient-change channel and coiled structure to achieve simultaneous impedance matching and minimal occupied space, leading to the ultra-thin thickness of the metasurface absorbers. Our conceived ultra-thin low-frequency broadband absorbers may lead to pragmatic implementations and applications in noise control field.

**Corresponding authors:**

*yifan.zhu@univ-lorraine.fr

*badreddine.assouar@univ-lorraine.fr

Sound absorption is important and even crucial in many occasions, such as architectural acoustics, reducing noise that can arise from machines, vehicle, or large-scale computers. The emergence of acoustic metamaterials [1-2] and acoustic metasurfaces [3] provide new ways to design acoustic functional devices. During the past several years, various acoustic metamaterial/metasurface-based absorbers [4] have been designed. Owing to the subwavelength feature of the acoustic resonant unit, the acoustic metamaterial/metasurface absorbers [5-21] can be very thin compared to conventional acoustic absorbers (such as porous and fibrous materials), which provide great flexibilities in real-world applications such as in architectural acoustics [22]. However, these designs based on coupled resonant systems [5-13] or coiling-up-space structures [12-19] usually have limited (less than 1/3 octave) or discontinuous bandwidths. Some designs have coupled more resonant units to improve the bandwidths to be about 2/3 octave [20-21].

To further improve the bandwidth (larger than 1 octave) of the acoustic absorber, various meta-structure designs are proposed, such as using fractal structure [23] or gradient structures [24]. However, the sample thicknesses in sound propagating direction [23-24] are comparable to the wavelength. Very recently, optimal sound-absorbing structures based on multiple resonant tubes [25] have been proposed to achieve highly efficient absorption within about 350Hz-3000Hz frequency range, with the structure thickness also close to $\lambda/10$ for the lowest working frequency. Another design [26] using gradient Helmholtz resonators has achieved bandwidth larger than 1 octave with the similar sample thickness ($\lambda/10$ for the lowest working frequency).

In this work, we propose a broadband acoustic metasurface absorber (AMA) having a deep-subwavelength thickness in the full working bandwidth [3]. Inspired by the optimal sound-absorbing structures reported in Ref. 25, but different from them, we propose to use a transformative coiled structure combined with a gradient-change channel as the unit cell to significantly decrease the thickness of the AMA [27]. It can achieve good impedance matching between the narrow coiled tubes and the free space, and at the same time minimize the occupied space. We propose two ultra-thin designs (*Sample* I/*Sample* II) 5.2cm/10.4cm, (*viz.*, $\lambda/15$ for the lowest working frequency, and $\lambda/7.5$ for the highest working frequency) for our designed AMAs. The obtained absorption is higher than 0.8 (averaged absorption is near 0.9) within the frequency ranges of 458Hz~968Hz/231Hz~491Hz, respectively.

The schematic diagram of the designed AMA is shown in Fig. 1. Figures 1(a) and 1(b) show the three-dimensional (3D) view and the top view of the AMA. The latter is a periodic array whose supercell (marked by red dashed frames) consists of 4×4 unit cells. Figure 1(c) shows the 4×4 unit cells denoted as number 1-16. We first show the design of *Sample* I with the total thickness of 5cm (considering the bottom thickness is 5.2cm). The side length of the square supercell is 2.4cm. The unit cells are designed in pairs as the blue dashed frames shows. A pair of unit cell is shown in Fig. 1(d) with the structural parameters marked. The unit cell is an air cavity consisting of two parts as follows:

(1) The gradient-change channel (GCC): the cross-sectional area of the channel is gradually changed. The thickness of the GCC part is 1cm.

(2) The coiled tube (CT): the coiled tube has a "gluttonous snake" (zigzag) structure. The thickness of the CT part is 4cm.

The GCC is employed to achieve good acoustic impedance matching between the opening of the air cavity (width is $w_0$=0.5cm) and the narrow CT (width is $w_1$=0.15cm). The impedance matching is necessary to make the incident acoustic wave totally coupled with the CT and dissipated inside. The thermal-viscous absorption can be very large (near to 1) at the Fabry–Perot (FP) resonant frequency of the CT. The resonant frequency is determined by the total length of the CT, *viz.*, the total length of CT corresponds to 1/4 of the resonant wavelength [25].

Another advantage of the deigned GCC is that it occupies minimum space. After using the GCC, $w_1=0.3w_0$ is obtained. In this case, obviously different from a simple straight tube (*viz.*, $w_0 \approx w_1$), we have more space to design the CT within the dimension of the unit cell, leading to the ultra-thin thickness of the AMA. In our design, we set 16 different total effective lengths ($L_{eff}$) of the CT to achieve 16 different resonant absorption peaks to ultimately realize broadband absorption [25]. $L_{eff}$ values are ranging from 7.55cm to 16.45cm for the 16 unit cells as shown in Fig. 1(e) (the curve marked by CT). The unit cells are designed in pairs as shown in Fig. 1(d-f), illustrating that the length sum of two CTs is fixed, making a pair of CTs occupy the entire space of 1.2cm×0.6cm×4cm. Considering the thickness of the GCC part, the effective total length of the 16 unit cells is ranging from 8.75cm to 19.3cm (the curve marked by GCC+CT). This is calculated by:

$$L_{eff} = \frac{f_r}{4c_0}, \qquad (1)$$

where $f_r$ is the FP resonant frequency of the air channel and $c_0$ is the sound speed in the air. It is noted that, the maximum of $L_{eff}$ is 19.3cm, which is realized in 5cm thickness, showing the high space utilization (386%) of the proposed method.

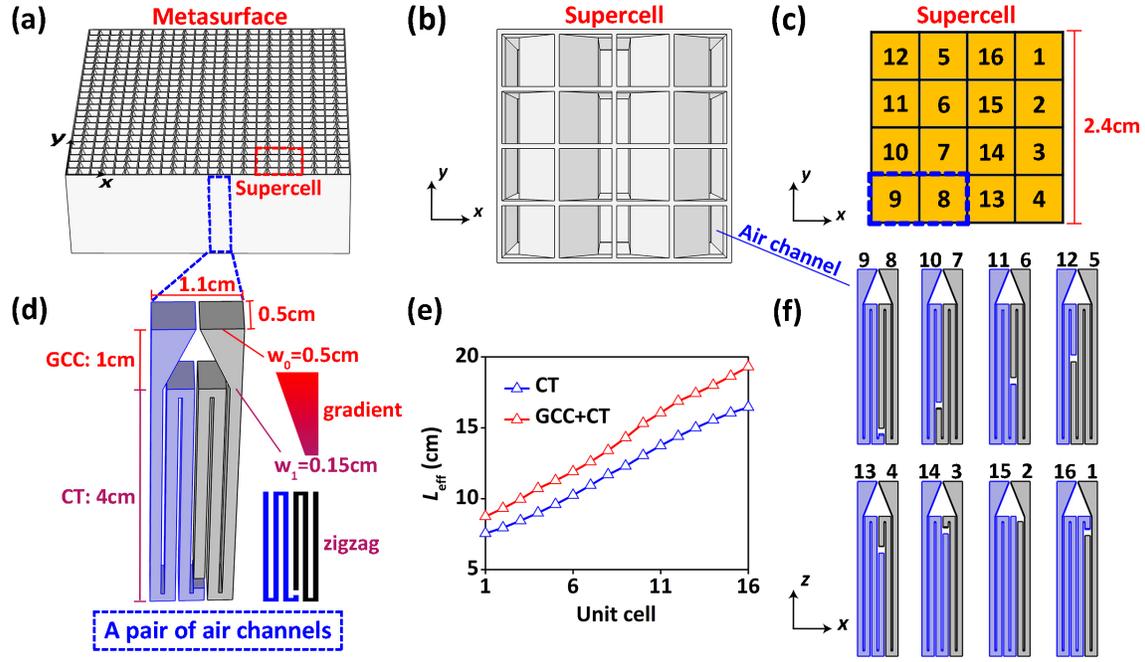

Fig. 1. The schematic diagram of the designed acoustic metasurface absorber (AMA). (a) The top view of the AMA. The supercell is marked by red dash lines. (b) The 3-dimensional view of the AMA. The supercell is marked by red dash lines. (c) The supercell consists of 4×4 unit cells denoted as 1-16. The side length of the supercell is 2.4cm. (d) Structural parameters of a pair of unit cells. The unit cell is an air cavity consisting of the gradient-change channel (GCC) (The widths are changing from 0.5cm to 0.15cm) and coiled tube (CT). (e) Effective lengths of CT and GCC+CT for 16 unit cells. (f) Structures of different unit cells (denoted as 1-16) with different coiled tube lengths.

We, then, carry out numerical and experimental analyzes of the proposed AMA. For the convenience of measurement, we have fabricated the *Sample* I with 2×2 supercells. The photograph of the sample is shown in Fig. 2(a) with the size of 5.4cm×5.4cm×5.2cm (Thickness is T=5.2cm). The size of the 2×2 supercells is 4.8cm×4.8cm×4.8cm. We have increased the outer wall thickness to make the sample mechanically stiffer). The sample is fabricated from polylactic acid (PLA) material via 3D printer with the sound speed of $c$=1200m/s, and the density of $\rho$=2700kg/m$^3$. The experimental setup is shown in Fig. 2(b). A lab-made acoustic impedance tube (inner size is 4.8cm×4.8cm), two Brüel & Kjær 1/4-inch-diameter microphones, and Brüel & Kjær measuring module "Acoustic Material Testing" are used to measure the absorption of the sample [28]. The numerical simulations are done by COMSOL Multiphysics 5.4a. The sound speed in air is set as $c_0$=343m/s, and the density of air is $\rho_0$=1.21kg/m$^3$.

The simulated and experimental results for the *Sample* I are shown in Fig. 2(c). The measured absorption is higher than 0.8 within 458Hz~968Hz (450Hz~1004Hz), showing good absorption effect in about one octave bandwidth. The average absorption within this bandwidth is 0.892 (0.911). The broadband absorption effect is attributed to the connection of 16 resonant frequencies (from 440Hz to 980Hz.) The resonant mode at 440Hz, 640Hz, 800Hz, and 980Hz are shown in the figure 2(c), which suggests that acoustic fields at these frequencies are localized in the unit cells 16, 8, 4, 1, respectively. The coupling between the unit cells in hybrid resonant regions lead to continuous high absorption spectrum. The results suggest that the thickness of the *sample* I is $\lambda$/15 for the lowest working frequency and is $\lambda$/7.5 for the highest working frequency (being deep-subwavelength for the full working bandwidth). For the full absorption, the equivalent acoustic impedance at the surface of the sample must be equal to the impedance of air, *viz.*, $Z=Z_0=\rho_0 c_0$. Here, the average absorption is about 0.9. The reason behind it could not reach to the full is that the GCC part still have little impedance mismatching which deviates the effective acoustic impedance from $\rho_0 c_0$. It is worth nothing that the simulated and measured sound absorptions out of the bandwidth (<450Hz, >1000Hz) is not close to zero. This is due to the fact that at the non-resonant (weak- resonant) region, sound wave is partly dissipative in the narrow air channels.

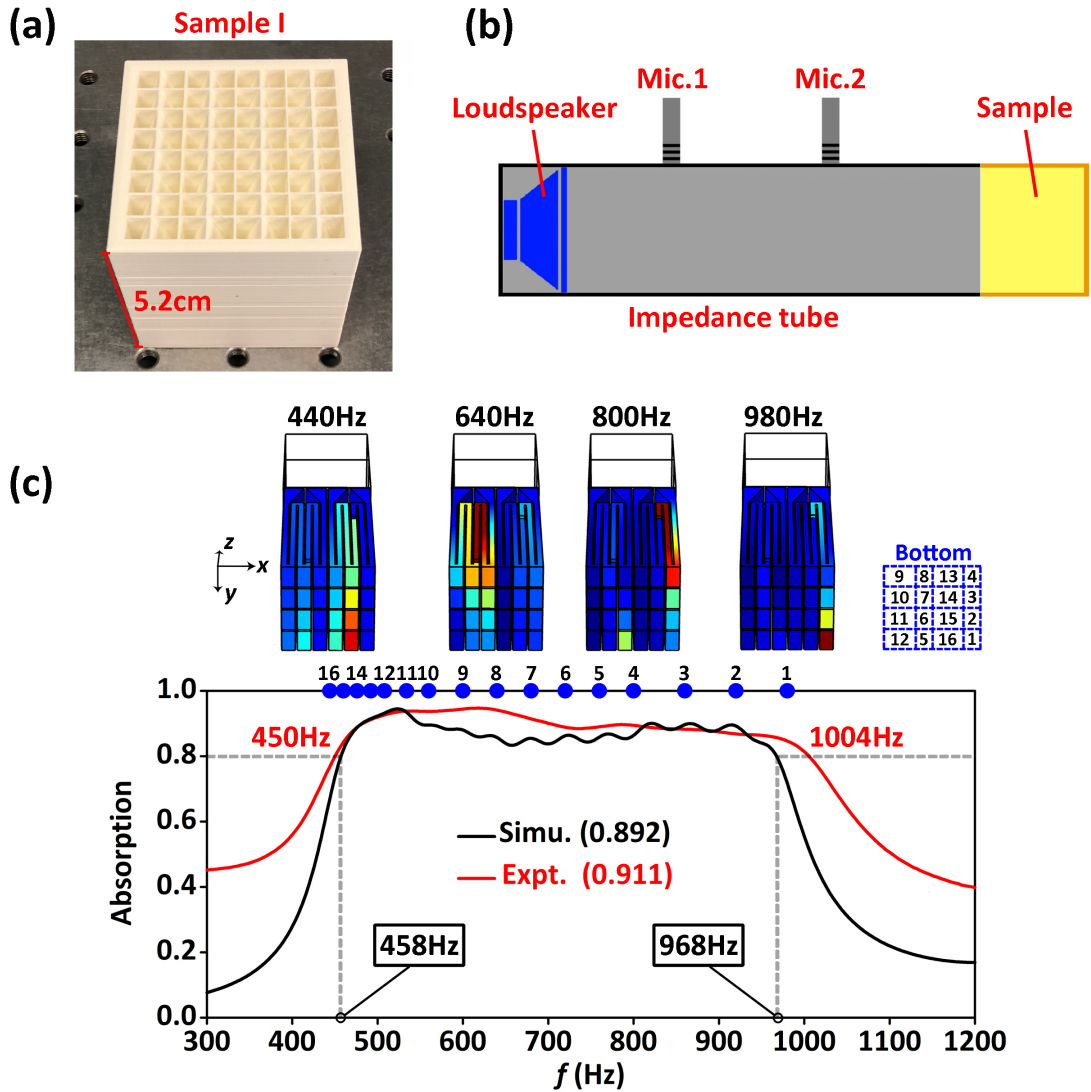

Fig. 2. (a) The photograph of *Sample* I with the size of 5.4cm×5.4cm×5.2cm. (b) Schematic of the sound absorption experiment using an impedance tube and two microphone technique. (c) The simulated and measured acoustic absorptions for the frequency range 300-1200Hz. The simulated and measured absorption is higher than 0.8 for the frequency range 458Hz~968Hz and 450Hz~1004Hz respectively. The simulated (measured) averaged absorption within this bandwidth is 0.892(0.911). The resonant mode at 440Hz, 640Hz, 800Hz, and 980Hz are shown for the corresponding resonant frequencies of the unit cell 16,8,4,1 respectively, corresponding to the resonant frequency of unit cell 16, 8, 4, 1, respectively.

We show another design with a lower frequency range absorption. Figure 3 shows the numerical and experimental results of *Sample* II. The supercell of *Sample* II (4.8cm×4.8cm×4.8cm) is enlarged proportionally from *Sample* I. *Sample* II is fabricated with one supercell with the size of 5.4cm×5.4cm×10.4cm (Thickness is T=10.4cm). Effective lengths $L_{eff}$ of CT and GCC+CT for 16 unit cells are shown in Fig. 3(b), which is designed for 16 resonant frequencies ranging from 220Hz to 490Hz. Figure 3(c) shows the simulated and the measured results of acoustic absorptions within 150Hz to 600Hz. The simulated (measured) absorption is larger than 0.8 within 231Hz~491Hz (232Hz~508Hz). The simulated (measured) averaged absorption within this bandwidth is 0.885 (0.897). Compared to the sample I, the absorption for the lower frequencies is slightly decreased for the sample II. This is can be numerically explained by the thickness of thermal-viscous boundary layer [29] which is given by:

$$d_{visc} = 0.22\text{mm} \times \sqrt{100/f} \qquad (2)$$

After normalized to the wavelength, we have achieved $d_{visc}/\lambda = 0.22\text{mm} \times \sqrt{100/\lambda c_0}$, indicating that the increase of $\lambda$ leads to decreasing of $d_{visc}/\lambda$. It indicates that the absorption at lower frequency is more difficult to achieve by the acoustic meta structure. We have also further studied the evanescent wave modes at the surface of the sample. Figure 3(d) shows the simulated and measured evanescent wave modes at the surface of the supercell for 300Hz and 400Hz, respectively. We have measured the acoustic fields in the openings of each air channel. The measured amplitude distributions of acoustic pressure agree well with the simulated ones, demonstrated the resonant mode at the surface of the sample.

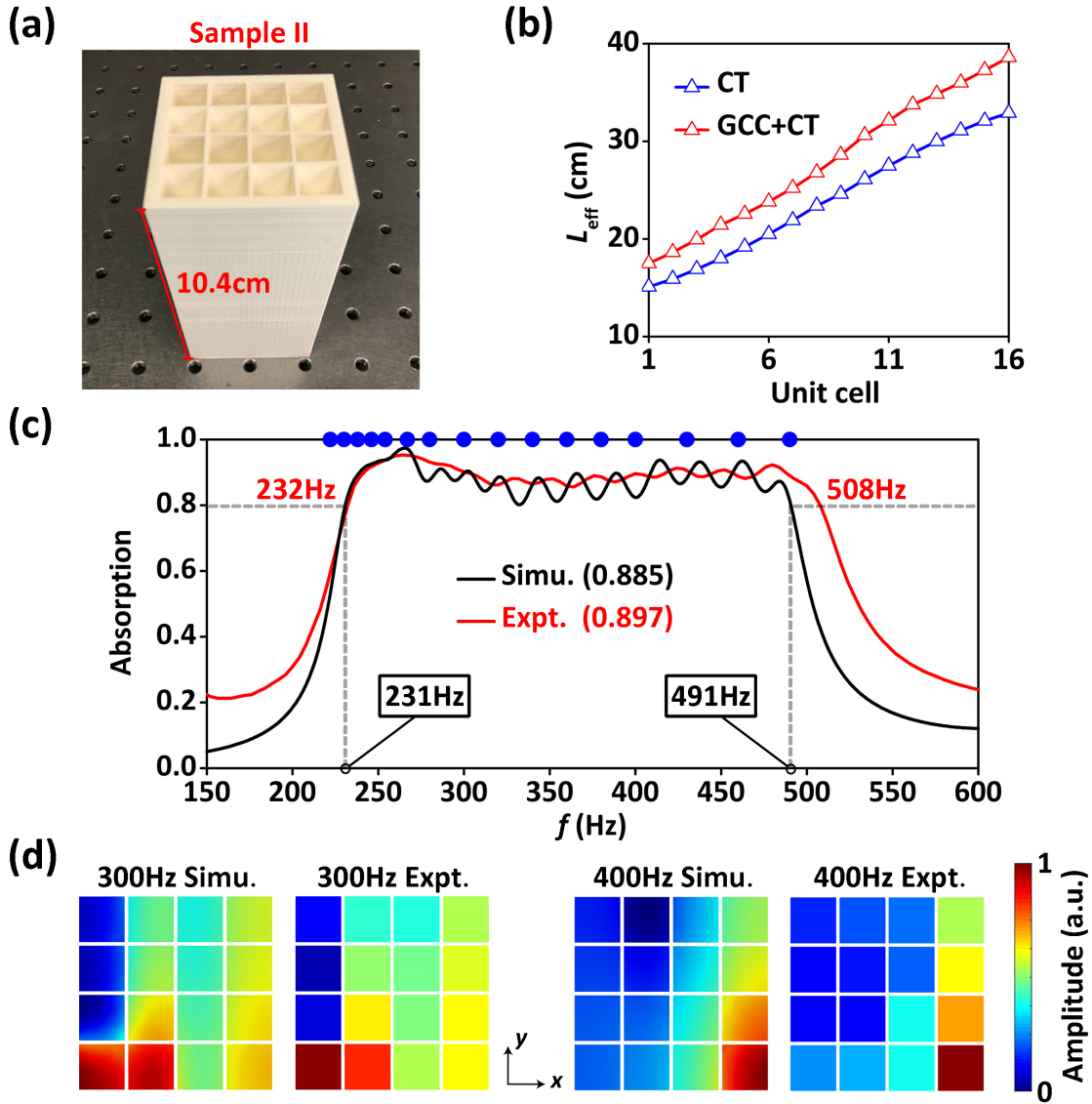

Fig. 3. (a) The photograph of *Sample* II with the size of 5.4cm×5.4cm×5.2cm. (b) Effective lengths of the CT and GCC+CT for 16 unit cells. (c) The simulated and measured acoustic absorptions for the frequency range 150Hz-600Hz. The simulated (measured) absorption is larger than 0.8 within 231Hz~491Hz (232Hz~508Hz). The simulated (measured) averaged absorption within this bandwidth is 0.885(0.897). (d) The simulated and measured evanescent wave modes at the surface of the sample for 300Hz and 400Hz, respectively.

In conclusion, we have designed and experimentally demonstrated the two designs of broadband low-frequency acoustic-metasurface-absorbers (larger than 1 octave), with the absorption larger than 0.8 (averaged absorption is near 0.9), and ultra-thin thickness ($\lambda/15$ for the lowest working frequency and $\lambda/7.5$ for the highest working frequency). The design strategy is based on combining gradient-change channel and coiled structure to achieve simultaneous good impedance matching and minimal occupied space. Specifically, the simulated (and measured) absorption of *Sample* I (thickness=5.2cm) is larger than 0.8 within 458Hz~968Hz (450Hz~1004Hz). The simulated (and measured) averaged absorption is 0.892 (0.911). The simulated (and measured) absorption of *Sample* II (thickness=10.4cm) is larger than 0.8 within 231Hz~491Hz (232Hz~508Hz). The simulated (and measured) averaged absorption is 0.885 (0.897). We can also design for other working wavelengths by scaling the sample size. Compared to the previous reported design of optimal sound-absorbing structures [25], the lowest working frequency can be reduced from about 350Hz to 231Hz, or say the thickness ($\lambda/15$) is 2/3 of the previous design ($\lambda/10$). On the other hand, our designs do not need to use the sponge at the surface of the sample [25] to ensure the continuous high absorption spectrum (larger than 0.8). Our proposed strategy may also be enlightening to reduce the thickness of acoustic functional meta-devices operating in the challenging low-frequency range. Eventually, the bandwidth of the proposed designs outperforms most of previous resonance-based AMAs, which could bring a real added value for the applications related to noise control and acoustic pollution suppression.

**Acknowledgments**

This work is supported by the Air Force Office of Scientific Research under award number FA9550-18-1-7021, and by la Région Grand Est and Institut Carnot ICEEL.